\documentclass[prd,twocolumn,eqsecnum,nofootinbib]{revtex4}
\usepackage{bm} 
\usepackage{amssymb}
\usepackage{graphicx} 
\newcommand{\base}[2]{e^{#1}_{#2}}
\newcommand{\EE}{\mbox{\,}^{\scriptscriptstyle \sf q\!}{\cal E}} 
\newcommand{\BB}{\mbox{\,}^{\scriptscriptstyle \sf q\!}{\cal B}} 
\newcommand{\jj}{\mbox{\,}^{\scriptscriptstyle \sf d\!}j}
\renewcommand{\SS}{\mbox{\,}^{\scriptscriptstyle \sf q\!}S}  
\begin{document}
\title{Light-cone coordinates based at a geodesic world line}
\author{Brent Preston and Eric Poisson}
\affiliation{Department of Physics, University of Guelph, Guelph, 
Ontario, Canada N1G 2W1}
\date{June 20, 2006} 
\begin{abstract} 
Continuing work initiated in an earlier publication [Phys.\ Rev.\ D 
{\bf 69}, 084007 (2004)], we construct a system of light-cone
coordinates based at a geodesic world line of an arbitrary curved   
spacetime. The construction involves (i) an advanced-time or a 
retarded-time coordinate that labels past or future light cones
centered on the world line, (ii) a radial coordinate that is an affine
parameter on the null generators of these light cones, and (iii)
angular coordinates that are constant on each generator. The spacetime
metric is calculated in the light-cone coordinates, and it is
expressed as an expansion in powers of the radial coordinate in terms
of the irreducible components of the Riemann tensor evaluated on the
world line. The formalism is illustrated in two simple applications,
the first involving a comoving world line of a spatially-flat
cosmology, the other featuring an observer placed on the axis of
symmetry of Melvin's magnetic universe.   
\end{abstract} 
\pacs{04.20.-q, 04.40.Nr} 
\maketitle

\section{Introduction} 

We continue a research program initiated in Ref.~\cite{poisson:04a},
which aims to construct and exploit light-cone coordinates based at an
arbitrary world line $\gamma$ of an arbitrary curved spacetime. The 
{\it retarded coordinates} introduced in Ref.~\cite{poisson:04a} are
denoted $(u,r,\theta,\phi)$ and are adapted to the {\it future light
cone} of each point $z$ on the selected world line. In this paper we
construct {\it advanced coordinates} $(v,r,\theta,\phi)$ which are
instead adapted to the {\it past light cone} of each point $z$ on the
world line; for simplicity we take the world line to be a geodesic of
the spacetime. (The advanced coordinates were introduced briefly in
Ref.~\cite{poisson:05}; this paper provides details that were not
given in the earlier work.) We collectively denote the light-cone
coordinates by $(w,r,\theta,\phi)$, with $w$ standing for either $u$ 
or $v$ depending on the context. In both cases the null coordinate $w$ 
is constant on each light cone, and it agrees with proper time $\tau$ 
at the cone's apex. The radial coordinate $r$ is an affine parameter
on the cone's null generators, and it measures the distance away
from the world line. The angular coordinates $\theta^A =
(\theta,\phi)$ are constant on each one of these generators. The
geometrical meaning of the light-cone coordinates is clear, and this
is one of their main virtues.    

The formalism developed in Ref.~\cite{poisson:04a} and in this paper
incorporates ideas formulated many years ago by Bondi and his  
collaborators \cite{bondi-etal:62, sachs:62}, and it complements a
line of research that was initiated by Synge \cite{synge:60} and
pursued by Ellis and his collaborators 
\cite{ellis-etal:85, ellis-etal:92a, ellis-etal:92b, ellis-etal:92c,
ellis-etal:92d, ellis-etal:94} in their work on observational
cosmology. While the central ideas exploited here are the same as with 
Synge and Ellis, our implementation is substantially different: While
Synge and Ellis sought definitions for their optical or observational 
coordinates that apply to large regions of the spacetime, our
considerations are limited to a small neighborhood of the world line. 

The introduction of retarded coordinates was motivated by the desire
to construct solutions to wave equations for massless fields that are
produced by a pointlike source moving on the world line. The retarded 
coordinates naturally incorporate the causal relation that exists
between the source and the field, and for this reason the solution
takes a simple explicit form (in the neighborhood in which the
coordinates are defined). The introduction of advanced coordinates is
motivated instead by the desire to construct solutions to the Einstein
field equations that describe black holes placed in a distribution of
matter or in a tidal environment. Such an application was described in
Ref.~\cite{poisson:05}, in which the metric of a tidally distorted
black hole was presented in advanced coordinates. In a companion paper
\cite{preston-poisson:06b} we use the guidance offered by the advanced
coordinates to formulate a light-cone gauge for black-hole
perturbation theory, and to calculate the metric of a black hole
immersed in a uniform magnetic field.    

A quasi-Cartesian version of the advanced coordinates is introduced
first in Sec.~II B, after reviewing some necessary geometrical
elements in Sec.~II A. The metric tensor in advanced coordinates is
constructed gradually in Secs.~II C through F, and its quasi-Cartesian
form is displayed in Eqs.~(\ref{2.6.3})--(\ref{2.6.5}). 

In Sec.~III A we combine the results obtained in Sec.~II with the
earlier results of Ref.~\cite{poisson:04a} and present the metric in a
general form suitable for both advanced and retarded coordinates. At
this stage the metric is expressed in terms of the Riemann tensor
evaluated on the world line $\gamma$. In Sec.~III B we begin to refine
the form of the metric by decomposing the Riemann tensor into its Weyl
and Ricci parts, and by further decompositing the Weyl and
energy-momentum tensors into their irreducible components. This leads
us, in Sec.~III C, to introduce {\it tidal and matter potentials} that
make the basic building blocks of the metric tensor. The potentials
are displayed in Table I, and the refined form of the metric is
displayed in Eqs.~(\ref{3.3.4})--(\ref{3.3.6}).  

In Sec.~IV we carry out a transformation of the metric from the 
quasi-Cartesian coordinates $\hat{x}^a$ to the quasi-spherical
coordinates $(r,\theta^A)$. The transformation, introduced in Sec.~IV
A, is the familiar one from flat spacetime: $\hat{x}^a = r
\Omega^a(\theta^A)$, or more explicitly, 
$\hat{x} = r\sin\theta\cos\phi$, 
$\hat{y} = r\sin\theta\sin\phi$, and 
$\hat{z} = r\cos\theta$. This final expression for the metric tensor,
in the coordinates $(w,r,\theta^A)$, is displayed in
Eqs.~(\ref{4.1.9})--(\ref{4.1.12}). It involves the angular components
of the tidal and matter potentials introduced in Sec.~III. As shown in
Table IV (with the results derived in Secs.~IV B and C), these are
naturally expressed as expansions in scalar, vector, and tensor
harmonics. The required spherical-harmonic functions are listed
in Tables II and III.   

In Sec.~V we present two simple applications of the light-cone
coordinates. In Sec.~V A we apply the formalism to a comoving world
line of a spatially-flat cosmology. In Sec.~V B we examine the metric
near the axis of symmetry of Melvin's magnetic universe
\cite{melvin:64, melvin:65, thorne:65}.  

Throughout the paper we work in geometrized units ($G=c=1$) and adhere
to the conventions of Misner, Thorne, and Wheeler \cite{MTW:73}. 

\section{Advanced coordinates} 

The presentation in this Section follows very closely Sec.~II of
Ref.~\cite{poisson:04a}. The material is very similar, but because of
important differences of sign that occur in various places, we present
here the details that are specific to the advanced coordinates. Other
details are omitted and can be obtained from Ref.~\cite{poisson:04a}.  

\subsection{Geometrical elements}

We first introduce some geometrical elements on the world line
$\gamma$ at which the advanced coordinates are based. The world line
is described by parametric relations $z^\mu(\tau)$ in which $\tau$
denotes proper time. Its normalized tangent vector is $u^\mu =
dz^\mu/d\tau$, and we assume that this satisfies the geodesic equation
$Du^\mu/d\tau = 0$. The world line is therefore a geodesic of the 
curved spacetime, and this assumption represents a loss of generality
relative to the construction of retarded coordinates presented in 
Ref.~\cite{poisson:04a}. While it would be a simple matter to restore 
this level of generality, we refrain from doing so in this work.      
Throughout we use Greek indices $\mu$, $\nu$, $\lambda$, $\rho$, etc.\
to refer to tensor fields defined, or evaluated, on the world line.     

We install on $\gamma$ an orthonormal tetrad that consists of the 
tangent vector $u^\mu$ and three spatial vectors $\base{\mu}{a}$. 
These are parallel transported on the world line, so that     
$D\base{\mu}{a}/d \tau = 0$. It is easy to check that this is
compatible with the requirement that the tetrad
$(u^\mu,\base{\mu}{a})$ be orthonormal everywhere on $\gamma$.  

From the tetrad on $\gamma$ we define a dual tetrad 
$(\base{0}{\mu},\base{a}{\mu})$ with the relations 
$\base{0}{\mu} = -u_\mu$ and $\base{a}{\mu} 
= \delta^{ab} g_{\mu\nu} \base{\nu}{b}$. The dual vectors
$\base{a}{\mu}$ also are parallel transported on the world line. The
tetrad and its dual give rise to the completeness relations    
\begin{eqnarray}
g^{\mu\nu} &=& -u^{\mu} u^{\nu} 
+ \delta^{ab} \base{\mu}{a} \base{\nu}{b}, 
\nonumber \\ 
& & \label{2.1.1} \\ 
g_{\mu\nu} &=& -\base{0}{\mu} \base{0}{\nu} 
+ \delta_{ab}\, \base{a}{\mu} \base{b}{\nu} 
\nonumber
\end{eqnarray} 
for the metric and its inverse evaluated on the world line.   

The advanced coordinates are constructed with the help of a null   
geodesic segment that links a given point $x$ to the world line. This
geodesic segment must be unique, and we thus restrict $x$ to be within
the normal convex neighborhood of $\gamma$. We denote by $\beta$ the
unique, future-directed null geodesic segment that goes from $x$ to
the world line, and $x' \equiv z(v)$ is $\beta$'s point of arrival on 
the world line; $v$ is the value of the proper-time parameter at this
point. To tensors at $x$ we assign the Greek indices $\alpha$,
$\beta$, $\gamma$, $\delta$, etc.; to tensors at $x'$ we assign the
indices $\alpha'$, $\beta'$, $\gamma'$, $\delta'$, and so on.      

From the tetrad $(u^{\alpha'},\base{\alpha'}{a})$ at $x'$ we obtain  
another tetrad $(\base{\alpha}{0},\base{\alpha}{a})$ at $x$ by
parallel transport on $\beta$. By raising the frame index and lowering
the vectorial index, we obtain also a dual tetrad at $x$:
$\base{0}{\alpha} = -g_{\alpha\beta} \base{\beta}{0}$ and 
$\base{a}{\alpha} = \delta^{ab} g_{\alpha\beta} \base{\beta}{b}$. The
metric at $x$ can then be expressed as 
\begin{equation}
g_{\alpha\beta} = -\base{0}{\alpha} \base{0}{\beta} + \delta_{ab} 
\base{a}{\alpha} \base{b}{\beta},   
\label{2.1.2}
\end{equation} 
and the parallel propagator \cite{synge:60} (also known as the
bivector of geodetic parallel displacement \cite{dewitt-brehme:60})
from $x'$ to $x$ is given by 
\begin{eqnarray} 
g^{\alpha}_{\ \alpha'}(x,x') &=& -\base{\alpha}{0} u_{\alpha'} 
+  \base{\alpha}{a} \base{a}{\alpha'},
\nonumber \\ 
& & \label{2.1.3} \\  
g^{\alpha'}_{\ \alpha}(x',x) &=& u^{\alpha'} \base{0}{\alpha} 
+ \base{\alpha'}{a} \base{a}{\alpha}. 
\nonumber 
\end{eqnarray}
This is defined such that if $A^\alpha$ is a vector that is parallel  
transported on $\beta$, then $A^\alpha(x) = g^\alpha_{\ \alpha'}(x,x') 
A^{\alpha'}(x')$ and $A^{\alpha'}(x') = g^{\alpha'}_{\ \alpha}(x',x) 
A^{\alpha}(x)$. Similarly, if $p_\alpha$ is a dual vector that is
parallel transported on $\beta$, then $p_\alpha(x) 
= g^{\alpha'}_{\ \alpha}(x',x) p_{\alpha'}(x')$ and $p_{\alpha'}(x')  
= g^{\alpha}_{\ \alpha'}(x,x') p_{\alpha}(x)$. 

The last ingredient we shall need is Synge's world function 
$\sigma(z,x)$ \cite{synge:60} (also known as the biscalar of geodetic
interval \cite{dewitt-brehme:60}). This is defined as half the squared 
geodesic distance between the world-line point $z(\tau)$ and a
neighboring point $x$. The derivative of the world function with
respect to $z^\mu$ is denoted $\sigma_\mu(z,x)$; this is a vector at
$z$ (and a scalar at $x$) that is known to be tangent to the geodesic
linking $z$ and $x$. The derivative of $\sigma(z,x)$ with respect to
$x^\alpha$ is denoted $\sigma_\alpha(z,x)$; this vector at $x$ (and
scalar at $z$) is also tangent to the geodesic. We
use a similar notation for multiple derivatives; for example,
$\sigma_{\mu\alpha} \equiv \nabla_\alpha \nabla_\mu \sigma$ and
$\sigma_{\alpha\beta} \equiv \nabla_\beta \nabla_\alpha \sigma$, where 
$\nabla_\alpha$ denotes a covariant derivative at $x$ while
$\nabla_\mu$ indicates covariant differentiation at $z$.    

The vector $-\sigma^\mu(z,x)$ can be thought of as a separation vector
between $x$ and $z$, pointing from the world line to $x$. When $x$ is
close to $\gamma$, $-\sigma^\mu(z,x)$ is small and can be used to
express bitensors in terms of ordinary tensors at
$z$ \cite{synge:60, dewitt-brehme:60}. For example, 
\begin{eqnarray} 
\sigma_{\mu\nu} &=& g_{\mu\nu} - \frac{1}{3} R_{\mu\lambda\nu\rho}
\sigma^\lambda \sigma^\rho + \cdots, 
\label{2.1.4} \\ 
\sigma_{\mu\alpha} &=& -g^\nu_{\ \alpha} \Bigl(g_{\mu\nu} 
+ \frac{1}{6} R_{\mu\lambda\nu\rho} \sigma^\lambda \sigma^\rho  
+ \cdots \Bigr),
\label{2.1.5} 
\end{eqnarray} 
where $g^\mu_{\ \alpha} \equiv g^\mu_{\ \alpha}(z,x)$ is the parallel
propagator and $R_{\mu\lambda\nu\rho}$ is the Riemann tensor evaluated
on the world line. 

\subsection{Definition of the advanced coordinates}  

In their quasi-Cartesian version, the retarded coordinates are defined
by     
\begin{equation}
\hat{x}^0 := v, \qquad
\hat{x}^a := -\base{a}{\alpha'}(x') \sigma^{\alpha'}(x,x'), \qquad  
\sigma(x,x') = 0.   
\label{2.2.1}
\end{equation}
The last statement indicates that $x' \equiv z(v)$ and $x$ are linked
by the null geodesic segment $\beta$, and we demand that this be
future-directed from $x$ to $x'$.  

From the fact that $\sigma^{\alpha'}$ is a null vector we obtain    
\begin{equation} 
r := (\delta_{ab} \hat{x}^a \hat{x}^b)^{1/2} 
= -u_{\alpha'} \sigma^{\alpha'}, 
\label{2.2.2}
\end{equation} 
and $r$ is a positive quantity because $\sigma^{\alpha'}$ is a
future-directed vector. We will see in Sec.~II C that $-r$ is an
affine parameter on $\beta$; this property adds credibility to the
idea that $r$ is meaningful measure of distance from $x$ to $x' \equiv
z(v)$.   

Another consequence of Eq.~(\ref{2.2.1}) is that 
\begin{equation}
\sigma^{\alpha'} = r \bigl( u^{\alpha'} - \Omega^a \base{\alpha'}{a}
\bigr), 
\label{2.2.3}
\end{equation}
where $\Omega^a := \hat{x}^a/r$ is a frame vector that satisfies  
$\delta_{ab} \Omega^a \Omega^b = 1$. 

A straightforward calculation reveals that under a displacement of the  
point $x$ (which induces a displacement of $x'$), the advanced
coordinates change according to   
\begin{eqnarray} 
d v &=& -l_\alpha\, d x^\alpha, 
\label{2.2.4} \\ 
d \hat{x}^a &=& -\base{a}{\alpha'} \sigma^{\alpha'}_{\ \beta'}
u^{\beta'}\, dv - \base{a}{\alpha'} 
\sigma^{\alpha'}_{\ \beta}\, d x^\beta,
 \label{2.2.5}
\end{eqnarray}
where $l_\alpha := -\sigma_{\alpha}/r$ is a future-directed null 
vector at $x$ that is tangent to $\beta$.  

\subsection{Advanced distance; null vector field}   

If we keep $x'$ linked to $x$ by the relation $\sigma(x,x') = 0$, then
\begin{equation}
r(x) = -\sigma_{\alpha'}(x,x') u^{\alpha'}(x') 
\label{2.3.1}
\end{equation}
can be viewed as an ordinary scalar field defined in a neighborhood 
of $\gamma$. We can compute the gradient of $r$ by finding how $r$ 
changes under a displacement of $x$ (which induces a displacement of
$x'$). The result is  
\begin{equation} 
\nabla_\beta r = \bigl( 
\sigma_{\alpha'\beta'} u^{\alpha'} u^{\beta'} \bigr) l_\beta 
- \sigma_{\alpha'\beta} u^{\alpha'}.
\label{2.3.2}
\end{equation} 

Similarly, we can view 
\begin{equation} 
l^{\alpha}(x) = -\frac{\sigma^{\alpha}(x,x')}{r(x)} 
\label{2.3.3}
\end{equation}
as an ordinary vector field, which is tangent to the congruence of
null geodesics that converge to $x'$. It is easy to check that
Eqs.~(\ref{2.3.2}) and (\ref{2.3.3}) imply   
\begin{equation}
l^\alpha \nabla_\alpha r = -1. 
\label{2.3.4}
\end{equation} 
In addition, combining the general statement $\sigma^{\alpha} =  
-g^{\alpha}_{\ \alpha'} \sigma^{\alpha'}$ with Eq.~(\ref{2.2.3}) gives 
\begin{equation} 
l^\alpha = g^{\alpha}_{\ \alpha'} \bigl( u^{\alpha'} - \Omega^a
\base{\alpha'}{a} \bigr); 
\label{2.3.5}
\end{equation} 
the vector at $x$ is therefore obtained by parallel transport of 
$u^{\alpha'} - \Omega^a \base{\alpha'}{a}$ on $\beta$. From this and
Eq.~(\ref{2.1.3}) we get the alternative expression 
\begin{equation} 
l^\alpha = \base{\alpha}{0} - \Omega^a \base{\alpha}{a},   
\label{2.3.6}
\end{equation} 
which confirms that $l^\alpha$ is a future-directed null vector field  
(recall that $\Omega^a = \hat{x}^a/r$ is a unit frame vector).    

The covariant derivative of $l_\alpha$ can be computed by finding  
how the vector changes under a displacement of $x$. This calculation
reveals that $l^\alpha$ satisfies the geodesic equation in 
affine-parameter form, $l^\beta \nabla_\beta l^\alpha = 0$, and 
Eq.~(\ref{2.3.4}) informs us that the affine parameter is in fact
$-r$. A displacement along a member of the congruence is therefore 
described by $dx^\alpha = -l^\alpha\, dr$. Specializing to the
advanced coordinates, and using Eqs.~(\ref{2.2.4}), (\ref{2.2.5}), and 
(\ref{2.3.3}), we find that this statement becomes $dv = 0$ and
$d\hat{x}^a = (\hat{x}^a/r)\, dr$, which integrate to $v =
\mbox{constant}$ and $\hat{x}^a = r \Omega^a$, respectively, with
$\Omega^a$ representing a constant unit vector. We have found that
the congruence of null geodesics that converge to $x'$ is described by     
\begin{equation}
v = \mbox{constant}, \qquad
\hat{x}^a = r \Omega^a(\theta^A)
\label{2.3.7}
\end{equation} 
in the advanced coordinates. Here, the two angles $\theta^A$ 
($A = 1, 2$) serve to parameterize the unit vector $\Omega^a$, which
is independent of $r$.       

Finally, we state without proof that $l^{\alpha}$ is hypersurface
orthogonal (the proof is contained in Ref.~\cite{poisson:04a}). This, 
together with the property that $l^\alpha$ satisfies the geodesic
equation in affine-parameter form, implies that there
exists a scalar field $v(x)$ such that   
\begin{equation}
l_\alpha = -\nabla_\alpha v. 
\label{2.3.8}
\end{equation} 
This scalar field was already identified in Eq.~(\ref{2.2.4}): it is  
numerically equal to the proper-time parameter of the world line at
$x'$. We conclude that the geodesics to which $l^\alpha$ is tangent
are the generators of the light cone $v = \mbox{constant}$. As 
Eq.~(\ref{2.3.7}) indicates, a specific generator is selected by 
choosing a direction $\Omega^a$ (which can be parameterized by two
angles $\theta^A$), and $-r$ is an affine parameter on each
generator. The geometrical meaning of the advanced coordinates is now 
completely clear; the construction is illustrated in Fig.~1.  

\begin{figure}
\includegraphics[angle=0,scale=0.4]{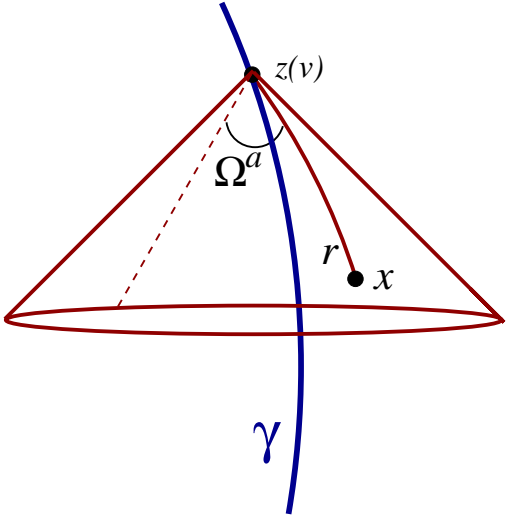} 
\caption{Advanced coordinates of a point $x$ relative to a world 
line $\gamma$. The advanced time $v$ selects a particular light cone,  
the unit vector $\Omega^a := \hat{x}^a/r$ selects a particular
generator of this light cone, and the advanced distance $r$ selects a   
particular point on this generator.} 
\end{figure} 

\subsection{Frame components of tensors on the world line}   

The metric at $x$ in the advanced coordinates will be expressed in  
terms of frame components of tensors evaluated on $\gamma$. We shall
need, in particular,  
\begin{eqnarray}
R_{a0b0}(v) &:=& R_{\alpha'\gamma'\beta'\delta'}\, \base{\alpha'}{a}  
u^{\gamma'} \base{\beta'}{b} u^{\delta'}, 
\nonumber \\ 
R_{abc0}(v) &:=& R_{\alpha'\gamma'\beta'\delta'}\, \base{\alpha'}{a} 
\base{\beta'}{b} \base{\gamma'}{c} u^{\delta'}, 
\label{2.4.1} \\  
R_{acbd}(v) &:=& R_{\alpha'\gamma'\beta'\delta'}\, \base{\alpha'}{a} 
\base{\gamma'}{c} \base{\beta'}{b} \base{\delta'}{d}.  
\nonumber 
\end{eqnarray}
These are the frame components of the Riemann tensor evaluated on  
$\gamma$; these quantities depend on $v$ only (recall that $v$ is
numerically equal to the proper-time parameter on the world line). 
We next form the useful combinations  
\begin{eqnarray}
P_{ab} &:=& R_{a0b0} - R_{acb0} \Omega^c 
- R_{bca0} \Omega^c + R_{acbd} \Omega^c \Omega^d 
\nonumber \\
&=& P_{ba}, 
\label{2.4.3} \\
P_{a} &:=& P_{ab}\Omega^b = R_{a0b0} \Omega^b - R_{abc0} 
\Omega^b \Omega^c, 
\label{2.4.4} \\
P &:=& P_{a} \Omega^a = R_{a0b0} \Omega^a \Omega^b, 
\label{2.4.5}
\end{eqnarray}
in which the quantities $\Omega^a := \hat{x}^a / r$ depend on the    
angles $\theta^A$ only --- they are independent of $v$ and $r$.  

\subsection{Coordinate displacements near $\gamma$} 

The changes in the quasi-Cartesian advanced coordinates under a
displacement of $x$ are given by Eqs.~(\ref{2.2.4}) and
(\ref{2.2.5}). In these we substitute the expansions for
$\sigma_{\alpha'\beta'}$ and $\sigma_{\alpha'\beta}$ that appear in
Eqs.~(\ref{2.1.4}) and (\ref{2.1.5}), as well as Eqs.~(\ref{2.2.3})
and (\ref{2.3.6}). After a straightforward calculation, we obtain the
following expressions for the coordinate displacements:   
\begin{eqnarray}
\hspace*{-20pt} 
dv &=& \bigl( \base{0}{\alpha}\, dx^\alpha \bigr) 
+ \Omega_a \bigl( \base{b}{\alpha}\, dx^\alpha \bigr),  
\label{2.5.1} \\ 
\hspace*{-20pt} 
d\hat{x}^a &=& \Bigl[ \frac{1}{2} r^2 P^a + O(r^3) \Bigr]
\bigl( \base{0}{\alpha}\, dx^\alpha \bigr) 
+ \Bigl[ \delta^a_{\ b} 
\nonumber \\ & & \mbox{} 
+ \frac{1}{6} r^2 
  ( P^a_{\ b} + 2 P^a \Omega_b ) + O(r^3) \Bigr] 
\bigl( \base{b}{\alpha}\, dx^\alpha \bigr).    
\label{2.5.2}
\end{eqnarray} 
Notice that the result for $dv$ is exact, but that $d\hat{x}^a$ is
expressed as an expansion in powers of $r$.  

\subsection{Metric near $\gamma$}   

It is straightforward to invert the relations of 
Eqs.~(\ref{2.5.1}) and (\ref{2.5.2}) and solve for $\base{0}{\alpha}\,  
dx^\alpha$ and $\base{a}{\alpha}\, dx^\alpha$. The results are  
\begin{eqnarray} 
\hspace*{-20pt}
\base{0}{\alpha}\, dx^\alpha &=& 
\Bigl[ 1 + \frac{1}{2} r^2 P + O(r^3) \Bigr]\, dv
- \Bigl[ \Omega_a 
\nonumber \\ & & \mbox{} 
- \frac{1}{6} r^2 ( P_a - P\Omega_a ) 
+ O(r^3) \Bigr]\, d\hat{x}^a, 
\label{2.6.1} \\ 
\hspace*{-20pt}
\base{a}{\alpha}\, dx^\alpha &=& 
\Bigl[ -\frac{1}{2} r^2 P^a + O(r^3) \Bigr]\, dv 
+ \Bigl[ \delta^a_{\ b} 
\nonumber \\ & & \mbox{} 
- \frac{1}{6} r^2 (P^a_{\ b} - P^a \Omega_b) 
+ O(r^3) \Bigr]\, d\hat{x}^b.
\label{2.6.2}
\end{eqnarray}  
These relations, when specialized to the advanced coordinates, give us   
the components of the dual tetrad $(\base{0}{\alpha},
\base{a}{\alpha})$ at $x$. The metric is then computed by involving
the completeness relations of Eq.~(\ref{2.1.1}). We find 
\begin{eqnarray}
g_{vv} &=& - 1 - r^2 P + O(r^3),  
\label{2.6.3} \\ 
g_{va} &=& \Omega_a - \frac{2}{3} r^2 
\bigl( P_a - P \Omega_a \bigr) + O(r^3),  
\label{2.6.4} \\
g_{ab} &=& \delta_{ab} - \Omega_a \Omega_b - \frac{1}{3} r^2 
\bigl( P_{ab} - P_a \Omega_b 
\nonumber \\ & & \mbox{} 
- \Omega_a P_b + P \Omega_a \Omega_b \bigr) + O(r^3).   
\label{2.6.5}
\end{eqnarray} 
We see that the metric possesses a directional ambiguity on the world
line: The metric at $r=0$ still depends on the vector $\Omega^a :=
\hat{x}^a/r$ that specifies the direction to the point $x$. The
advanced coordinates are therefore singular on the world line, and 
tensor components cannot be defined on $\gamma$. This poses no
particular difficulty because we can always work, as we have been 
doing, with {\it frame components} of tensors instead of tensorial
components.  

\section{Light-cone coordinates; decomposition of the Riemann tensor}  

\subsection{Retarded and advanced coordinates} 

The developments of Sec.~II parallel very closely the construction of
retarded coordinates described in Ref.~\cite{poisson:04a}. The combined
set of results is a coordinate system $(w,\hat{x}^a)$ that refers
either to past light cones (advanced coordinates, $w \equiv v$, $\eta
\equiv +1$) or to future light cones (retarded coordinates, $w \equiv
u$, $\eta \equiv -1$) centered on a geodesic world line. In either case
the metric is expressed as 
\begin{eqnarray}
g_{ww} &=& - 1 - r^2 P + O(r^3),  
\label{3.1.1} \\ 
\eta g_{wa} &=& \Omega_a - \frac{2}{3} r^2 
\bigl( P_a - P \Omega_a \bigr) + O(r^3),  
\label{3.1.2} \\
g_{ab} &=& \delta_{ab} - \Omega_a \Omega_b - \frac{1}{3} r^2 
\bigl( P_{ab} - P_a \Omega_b 
\nonumber \\ & & \mbox{} 
- \Omega_a P_b + P \Omega_a \Omega_b \bigr) + O(r^3),    
\label{3.1.3}
\end{eqnarray} 
with 
\begin{eqnarray}
P_{ab} &:=& R_{a0b0} - \eta R_{acb0} \Omega^c 
- \eta R_{bca0} \Omega^c 
\nonumber \\ & & \mbox{} 
+ R_{acbd} \Omega^c \Omega^d 
\label{3.1.4} \\
P_{a} &:=& R_{a0b0} \Omega^b - \eta R_{abc0} \Omega^b \Omega^c,  
\label{3.1.5} \\
P &:=& R_{a0b0} \Omega^a \Omega^b.  
\label{3.1.6}
\end{eqnarray}
The Riemann tensor is evaluated at the advanced/retarded point $x'
\equiv z(w)$, and its frame components are defined as in
Eqs.~(\ref{2.4.1}). Our subsequent developments will apply to the
general metric of Eqs.~(\ref{3.1.1})--(\ref{3.1.3}). They will not
distinguish between the advanced and retarded coordinates, and we will
collectively refer to them as {\it light-cone coordinates}. As was
stated previously, $w$ stands for either $v$ or $u$, and $\eta$  
is an indicator that takes the value $+1$ for the advanced
coordinates, and the value $-1$ for the retarded coordinates.  

\subsection{Decomposition of the Riemann tensor} 

To bring the metric to a more explicit form we decompose the Riemann
tensor into its Weyl and Ricci parts, and we involve the Einstein
field equations to relate the Ricci tensor to the energy-momentum
tensor of the matter distribution. This gives  
\begin{eqnarray} 
R_{\alpha'\beta'\gamma'\delta'} &=& C_{\alpha'\beta'\gamma'\delta'}  
+ 8\pi \bigl( g_{\alpha'[\gamma'} T_{\delta']\beta'} 
- g_{\beta'[\gamma'} T_{\delta']\alpha'} \bigr) 
\nonumber \\ & & \mbox{} 
- \frac{16\pi}{3} g_{\alpha'[\gamma'} g_{\delta']\beta'}\, 
\mbox{}^4 T,   
\label{3.2.1}
\end{eqnarray} 
where $C_{\alpha'\beta'\gamma'\delta'}$ is the Weyl tensor,
$T_{\alpha'\beta'}$ is the energy-momentum tensor, $\mbox{}^4 T :=
T^{\alpha'}_{\ \alpha'}$ is its four-dimensional trace, and where the
square brackets indicate antisymmetrization of the enclosed indices. 

We next project the Weyl tensor onto the tetrad  
$(u^{\alpha'},\base{\alpha'}{a})$ and decompose the projections into  
irreducible components, according to \cite{thorne-hartle:85}   
\begin{eqnarray} 
C_{a0b0} &=& {\cal E}_{ab}, 
\nonumber \\ 
C_{abc0} &=& \varepsilon_{abp} {\cal B}^p_{\ c}, 
\label{3.2.2} \\ 
C_{abcd} &=& \delta_{ac} {\cal E}_{bd} - \delta_{bc} {\cal E}_{ad} 
- \delta_{ad} {\cal E}_{bc} + \delta_{bd} {\cal E}_{ac},  
\nonumber 
\end{eqnarray} 
where $\varepsilon_{abc}$ is the flat-space permutation symbol. The 
electric components of the Weyl tensor are denoted ${\cal E}_{ab}$,
while the magnetic components are denoted ${\cal B}_{ab}$. These
tensors are symmetric and tracefree, so that, for example, 
${\cal E}_{ba} = {\cal E}_{ab}$ and ${\cal B}^a_{\ a} = 0$. Because
the Weyl tensor is evaluated on the world line, ${\cal E}_{ab}$ and
${\cal B}_{ab}$ are functions of the null coordinate $w$ only.   

We perform similar operations on the energy-momentum tensor, and
introduce the notation   
\begin{eqnarray} 
T_{00} &=& \rho, 
\nonumber \\ 
T_{0a} &=& -j_a, 
\label{3.2.3} \\ 
T_{ab} &=& S_{ab} + \frac{1}{3} \delta_{ab} T, 
\nonumber 
\end{eqnarray} 
where $S_{ab}$ is symmetric and tracefree, and $T := \delta^{ab} 
T_{a b} = T_{\alpha'\beta'} (\delta^{ab} \base{\alpha'}{a}
\base{\beta'}{b}) = T_{\alpha'\beta'} (g^{\alpha'\beta'} + u^{\alpha'}
u^{\beta'}) = \mbox{}^4 T + \rho$ is the spatial trace of the
energy-momentum tensor. The quantity $\rho$ represents the mass-energy 
density measured by an observer moving on the world line $\gamma$,
$j_a$ is the flux of mass-energy traveling in the direction of the 
base vector $\base{\alpha'}{a}$, $S_{ab}$ is the tracefree part of the
stress tensor, and $\frac{1}{3} T$ is an isotropic pressure. These
quantities also are functions of $w$ only.      

Substituting Eqs.~(\ref{3.2.2}) and (\ref{3.2.3}) into
Eq.~(\ref{3.2.1}), and then this into Eq.~(\ref{3.1.4}), produces a
decomposition of $P_{ab}$ into its irreducible pieces. We obtain  
\begin{eqnarray} 
P_{ab} &=& 2 {\cal E}_{ab} 
- \Omega_{a} {\cal E}_{bc} \Omega^c
- \Omega_{b} {\cal E}_{ac} \Omega^c 
+ \delta_{ab} {\cal E}_{cd} \Omega^c \Omega^d
\nonumber \\ & & \mbox{} 
- \eta\bigl( \varepsilon_{apq} \Omega^p {\cal B}^p_{\ b}
           + \varepsilon_{bpq} \Omega^p {\cal B}^p_{\ a} \bigr) 
\nonumber \\ & & \mbox{} 
+ \frac{4\pi}{3} \bigl( 3\delta_{ab} - 2\Omega_a \Omega_b \bigr) \rho 
\nonumber \\ & & \mbox{} 
- 4\pi\eta \bigl( j_{a} \Omega_b + j_b \Omega_a 
                - 2\delta_{ab} j_c \Omega^c \bigr) 
\nonumber \\ & & \mbox{} 
-4\pi \bigr( \Omega_{a} S_{bc} \Omega^c
+ \Omega_{b} S_{ac} \Omega^c 
- \delta_{ab} S_{cd} \Omega^c \Omega^d \bigr) 
\nonumber \\ & & \mbox{} 
+ \frac{4\pi}{3} \delta_{ab} T. 
\label{3.2.4} 
\end{eqnarray} 

\subsection{Tidal and matter potentials} 

At this stage it is useful to involve the irreducible quantities 
${\cal E}_{ab}(w)$ and ${\cal B}_{ab}(w)$ in the definition of a
number of {\it tidal potentials}. We also involve $j_a(w)$ and
$S_{ab}(w)$ in the definition of {\it matter potentials}. These
potentials, which are displayed in Table I, form the elementary
building blocks of the metric tensor. Each potential is identified by
its multipole content. For example, 
$\EE := {\cal E}_{ab} \Omega^a \Omega^b$ is a quadrupolar potential by
virtue of the fact that ${\cal E}_{ab}$ is symmetric and tracefree. As
another example, $\jj := j_a \Omega^a$ is a dipolar potential. Table I
also introduces vectorial and tensorial potentials that possess the
property of being transverse, meaning that each vector or tensor is
orthogonal to the unit frame vector $\Omega^a$. Finally, the tensor
potentials $\EE_{ab}$ and $\BB_{ab}$ have the additional property of
being symmetric and tracefree.       

\begin{table}
\caption{Tidal and matter potentials. Each potential is identified
with a sans-serif superscript that specifies its multipole content. A
potential labeled with a ``{\sf q}'' is a quadrupole field, and one
labeled with an ``{\sf d}'' is a dipole field. The vector and tensor
potentials are all orthogonal to $\Omega^a$, and all tensors are
symmetric and tracefree.}  
\begin{ruledtabular}
\begin{tabular}{l}
$\EE = {\cal E}_{cd} \Omega^c \Omega^d$ \\ 
$\EE_a = (\delta_a^{\ c} - \Omega_a \Omega^c) 
    {\cal E}_{cd} \Omega^d$ \\
$\EE_{ab} = 2(\delta_a^{\ c} - \Omega_a \Omega^c)(\delta_b^{\ d} 
    -\Omega_b \Omega^d) {\cal E}_{cd} 
    + (\delta_{ab} - \Omega_a \Omega_b) \EE$ \\ 
$\BB_a = \varepsilon_{apq} \Omega^p {\cal B}^q_{\ c} \Omega^c$ \\ 
$\BB_{ab} = \varepsilon_{apq} \Omega^p {\cal B}^q_{\ c} 
    (\delta^c_{\ b} - \Omega^c \Omega_b) 
+ \varepsilon_{bpq} \Omega^p {\cal B}^q_{\ c} 
    (\delta^c_{\ b} - \Omega^c \Omega_a) $ \\  
$\jj = j_c \Omega^c$ \\ 
$\jj_a = (\delta_a^{\ c} - \Omega_a \Omega^c) j_c$ \\ 
$\SS = S_{cd} \Omega^c \Omega^d$ \\ 
$\SS_a = (\delta_a^{\ c} - \Omega_a \Omega^c) S_{cd} \Omega^d$ 
\end{tabular}
\end{ruledtabular}
\end{table}   

It is easy to check that $P_{ab}$ is expressed in terms of the
tidal and matter potentials as 
\begin{eqnarray} 
P_{ab} &=& \EE_{ab} + 2\Omega_{(a} \EE_{b)} + \Omega_a \Omega_b \EE 
\nonumber \\ & & \mbox{} 
-\eta \bigl( \BB_{ab} + 2 \Omega_{(a} \BB_{b)} \bigr) 
\nonumber \\ & & \mbox{} 
+ 4\pi \bigl(\delta_{ab} - \Omega_a \Omega_b \bigr) \rho 
+ \frac{4\pi}{3} \Omega_a \Omega_b \rho 
\nonumber \\ & & \mbox{} 
+ 8\pi \eta \bigl[ (\delta_{ab} - \Omega_{ab}) \jj 
- \Omega_{(a} \jj_{b)} \bigr] 
\nonumber \\ & & \mbox{} 
+ 4\pi \bigl(\delta_{ab} - \Omega_a \Omega_b \bigr) \SS 
- 8\pi \Omega_{(a} \SS_{b)} - 4\pi \Omega_a \Omega_b \SS 
\nonumber \\ & & \mbox{} 
+ \frac{4\pi}{3} \bigl(\delta_{ab} - \Omega_a \Omega_b \bigr) T 
+ \frac{4\pi}{3} \Omega_a \Omega_b T. 
\label{3.3.1}
\end{eqnarray} 
We observe that $P_{ab}$ is now decomposed into transverse-transverse
components that are fully orthogonal to $\Omega_a$,
transverse-longitudinal components that are partly orthogonal to and 
partly aligned with $\Omega_a$, and longitudinal-longitudinal
components that are proportional to $\Omega_a \Omega_b$. Contracting
Eq.~(\ref{3.3.1}) with $\Omega^b$ produces 
\begin{eqnarray} 
P_a &=& \EE_a + \Omega_a \EE - \eta \BB_a 
+ \frac{4\pi}{3} \Omega_a \rho  - 4\pi \eta \jj_a 
\nonumber \\ & & \mbox{} 
- 4\pi \SS_a - 4\pi \Omega_a \SS + \frac{4\pi}{3} \Omega_a T, 
\label{3.3.2}
\end{eqnarray}
and contracting this with $\Omega^a$ gives 
\begin{equation}
P = \EE + \frac{4\pi}{3} \rho - 4\pi \SS + \frac{4\pi}{3} T. 
\label{3.3.3}
\end{equation} 

Substituting Eqs.~(\ref{3.3.1})--(\ref{3.3.3}) into
Eqs.~(\ref{3.1.1})--(\ref{3.1.3}) produces our final expression for 
the metric tensor in the quasi-Cartesian version of the light-cone 
coordinates. We obtain, after simplification, 
\begin{eqnarray} 
\hspace*{-20pt} 
g_{ww} &=& -1 - r^2 \EE - \frac{4\pi}{3} r^2 
\bigl( \rho - 3 \SS + T \bigr) + O(r^3), 
\label{3.3.4} \\ 
\hspace*{-20pt} 
\eta g_{wa} &=& \Omega_a - \frac{2}{3} r^2 \bigl( \EE_a 
- \eta \BB_a \bigr) 
\nonumber \\ & & \mbox{} 
+ \frac{8\pi}{3} \eta r^2 \bigl( \SS_a 
+ \eta \jj_a \bigr) + O(r^3), 
\label{3.3.5} \\ 
\hspace*{-20pt} 
g_{ab} &=& \delta_{ab} - \Omega_a \Omega_b - \frac{1}{3} r^2 \bigl(
\EE_{ab} - \eta \BB_{ab} \bigr) 
\nonumber \\ & & \mbox{} 
- \frac{4\pi}{3} r^2 \bigl( 
\delta_{ab} - \Omega_a \Omega_b \bigr) \bigl( \rho + 2\eta \jj + \SS 
+ {\textstyle \frac{1}{3}} T \bigr) 
\nonumber \\ & & \mbox{} 
+ O(r^3). 
\label{3.3.6}
\end{eqnarray} 
We observe that the metric is neatly expressed in terms of the
monopolar ``potentials'' $\rho$ and $T$, the dipolar potentials $\jj$
and $\jj_a$, the quadrupolar tidal potentials $\EE$, $\EE_a$,
$\EE_{ab}$, $\BB_a$, $\BB_{ab}$, and the quadrupolar matter potentials
$\SS$ and $\SS_a$. We observe also that $g_{wa}$ contains both
longitudinal and transverse pieces, while $g_{ab}$ is fully
transverse. We recall that the coordinates are advanced when $\eta =
+1$ (then $w \equiv v$) and that they are retarded when $\eta = -1$
(then $w \equiv u$).    

\section{Angular coordinates} 

\subsection{Transformation to angular coordinates} 

Because the frame vector $\Omega^a := \hat{x}^a/r$ satisfies 
$\delta_{ab} \Omega^a \Omega^b = 1$, it can be parameterized by two  
angles $\theta^A$. A canonical choice for the parameterization is
\begin{equation} 
\Omega^a = (\sin\theta \cos\phi, \sin\theta \sin\phi, \cos\theta). 
\label{4.1.1}
\end{equation} 
It is then convenient to perform a coordinate transformation from
$\hat{x}^a$ to $(r,\theta^A)$ using the relations $\hat{x}^a = r
\Omega^a(\theta^A)$. (Recall from Sec.~II C that the angles $\theta^A$
are constant on the generators of the light cones $w =
\mbox{constant}$, and that $\pm r$ is an affine parameter on these
generators. The relations $\hat{x}^a = r \Omega^a$ therefore describe
the behavior of the generators.) The differential form of the
coordinate transformation is      
\begin{equation}
d\hat{x}^a = \Omega^a\, dr + r \Omega^a_A\, d\theta^A, 
\label{4.1.2}
\end{equation}
where the transformation matrix 
\begin{equation} 
\Omega^a_A := \frac{\partial \Omega^a}{\partial \theta^A}
\label{4.1.3}
\end{equation} 
satisfies the identity $\Omega_a \Omega^a_A = 0$. 

We introduce the quantities 
\begin{equation}
\Omega_{AB} := \delta_{ab} \Omega^a_A \Omega^b_B, 
\label{4.1.4}
\end{equation}
which act as a (nonphysical) metric on the submanifold spanned
by the angular coordinates. In the canonical parameterization of
Eq.~(\ref{4.1.1}), $\Omega_{AB} = \mbox{diag}(1,\sin^2\theta)$, and
the metric is that of a round two-sphere of unit radius. We use the
inverse of $\Omega_{AB}$, denoted $\Omega^{AB}$, to raise upper-case
Latin indices. We then define the new object  
\begin{equation}
\Omega^A_a := \delta_{ab} \Omega^{AB} \Omega^b_B 
\label{4.1.5}
\end{equation} 
which satisfies the identities 
\begin{equation}
\Omega^A_a \Omega^a_B = \delta^A_B, \qquad
\Omega^a_A \Omega^A_b = \delta^a_{\ b} - \Omega^a \Omega_b. 
\label{4.1.6}
\end{equation} 
The first result is a direct consequence of the definition, and the
second result follows from the fact that both sides are symmetric in
$a$ and $b$, orthogonal to $\Omega_a$ and $\Omega^b$, and have the
same trace.   

The Levi-Civita tensor on $S^2$ is constructed as 
\begin{equation}
\varepsilon_{AB} := \varepsilon_{abc} \Omega^a_A \Omega^b_B \Omega^c, 
\label{4.1.7}
\end{equation}
where $\varepsilon_{abc}$ is the Cartesian permutation symbol; in 
the canonical coordinates we have $\varepsilon_{\theta\phi} =
\sin\theta$. 

We let $D_A$ denote the covariant derivative operator compatible with
$\Omega_{AB}$, so that $D_A \Omega_{BC} = 0$. It is easy to show that
$D_A \varepsilon_{BC} = 0$ and 
\begin{equation} 
\Omega^a_{AB} := D_B \Omega^a_{A} = -\Omega^a \Omega_{AB}. 
\label{4.1.8} 
\end{equation} 

When we apply the coordinate transformation of Eq.~(\ref{4.1.2}) to
the metric of Eqs.~(\ref{3.3.4})--(\ref{3.3.6}) we find that the only 
nonvanishing components of the metric tensor are now given by  
\begin{eqnarray} 
\hspace*{-15pt} 
g_{ww} &=& -1 - r^2 \EE - \frac{4\pi}{3} r^2 
\bigl( \rho - 3 \SS + T \bigr) + O(r^3), 
\label{4.1.9} \\ 
\hspace*{-15pt} 
g_{wr} &=& \eta, 
\label{4.1.10} \\ 
\hspace*{-15pt} 
g_{wA} &=& -\frac{2}{3} \eta r^3 \bigl( \EE_A  
- \eta \BB_A \bigr) 
\nonumber \\ & & \mbox{} 
+ \frac{8\pi}{3} \eta r^3 \bigl( \SS_A + \eta \jj_A \bigr) + O(r^4),  
\label{4.1.11} \\ 
\hspace*{-15pt} 
g_{AB} &=& r^2 \Omega_{AB} - \frac{1}{3} r^4 \bigl(
\EE_{AB} - \eta \BB_{AB} \bigr) 
\nonumber \\ & & \mbox{} 
- \frac{4\pi}{3} r^4 \Omega_{AB} 
\bigl( \rho + 2\eta \jj + \SS + {\textstyle \frac{1}{3}} T \bigr) 
\nonumber \\ & & \mbox{} 
+ O(r^5),  
\label{4.1.12}
\end{eqnarray} 
where $\EE_A := \EE_a \Omega^a_A$, $\EE_{AB} := \EE_{ab} \Omega^a_A
\Omega^b_B$, and so on. The results $g_{wr} = \eta$, $g_{rr} = 0$, and
$g_{rA} = 0$ are exact, and they follow from the light-cone nature of
the coordinates. [For example, for the advanced coordinates we have 
$l_\alpha = (-1,0,0,0)$ and $l^\alpha = (0,-1,0,0)$, where we use the
ordering $(v,r,\theta,\phi)$; these relations imply that $g_{vr} = 1$
and $g_{rr} = g_{r\theta} = g_{r\phi} = 0$.] Once more we recall that
the coordinates are advanced when $\eta = +1$ (then $w \equiv v$), and 
that they are retarded when $\eta = -1$ (then $w \equiv u$). 

\subsection{Tidal and matter potentials in spherical coordinates} 

According to Table I, the tidal potential $\EE$ is defined by $\EE =
{\cal E}_{cd} \Omega^c \Omega^d$. Differentiating this with respect to 
$\theta^A$ gives $D_A \EE = 2 \Omega^c_A {\cal E}_{cd} \Omega^d$. In
view of the identity $\Omega_c \Omega^c_A = 0$ we may write this as  
$D_A \EE = 2\Omega^a_A (\delta_a^{\ c} - \Omega_a \Omega^c) 
{\cal E}_{cd} \Omega^d$. Referring once more to Table I, we see that
this is $D_A \EE = 2\Omega^a_A \EE_a$ and we conclude that 
\begin{equation} 
\EE_A = \frac{1}{2} D_A \EE. 
\label{4.2.1}
\end{equation} 
Acting on $\EE$ with two derivative operators gives $D_A D_B \EE  
= 2 {\cal E}_{cd} \Omega^c_A \Omega^d_B + 2 {\cal E}_{cd} \Omega^c
\Omega^d_{AB}$. Using Eq.~(\ref{4.1.8}) produces $D_A D_B \EE =    
2 {\cal E}_{cd} \Omega^c_A \Omega^d_B - 2 {\cal E}_{cd} \Omega^c
\Omega^d \Omega_{AB} = 2 {\cal E}_{cd} \Omega^c_A \Omega^d_B 
- 2 \Omega_{AB} \EE$. We write this in the form $(D_A D_B + 3
\Omega_{AB}) \EE = 2 {\cal E}_{cd} \Omega^c_A \Omega^d_B 
+ \Omega_{AB} \EE = 2 {\cal E}_{cd} \Omega^a_A \Omega^b_B 
(\delta_a^{\ c} - \Omega_a \Omega^c) (\delta_b^{\ d} 
- \Omega_b \Omega^d) {\cal E}_{cd} + \Omega^a_A \Omega^b_B
(\delta_{ab} - \Omega_a \Omega_b) \EE$, after involving
Eq.~(\ref{4.1.4}). Consulting Table I once more, we see that the
right-hand side is equal to $\Omega^a_A \Omega^b_B \EE_{ab} =:
\EE_{AB}$ and we conclude that 
\begin{equation} 
\EE_{AB} = \bigl(D_A D_B + 3 \Omega_{AB} \bigr) \EE. 
\label{4.2.2}
\end{equation} 
We observe that this tensor is tracefree, because $\Omega^{AB}
\EE_{AB} = \Omega^{AB} \Omega^a_A \Omega^b_B \EE_{ab} = (\delta^{ab} -
\Omega^a \Omega^b) \EE_{ab} = 0$, after involving Eq.~(\ref{4.1.6})
and the fact that $\EE_{ab}$ is transverse and tracefree (in the
Cartesian sense). The equation $(\Omega^{AB} D_A D_B + 6) \EE = 0$,
which we obtain from Eq.~(\ref{4.2.2}), reveals that $\EE(w,\theta^A)$ 
is a spherical-harmonic function of degree $l = 2$. 

We define a magnetic potential $\BB := {\cal B}_{cd} \Omega^c
\Omega^d$ and differentiate it with respect to $\theta^B$, giving 
$D_B \BB = 2\Omega^c_B {\cal B}_{cd} \Omega^d$. We next multiply this
by the Levi-Civita tensor of Eq.~(\ref{4.1.7}) and get
$-\varepsilon_A^{\ B} D_B \BB = -2 \varepsilon_{apq} \Omega^a_A
\Omega^{pB} \Omega^q \Omega^c_B {\cal B}_{cd} \Omega^d$. Using
Eq.~(\ref{4.1.6}) and the antisymmetry property of the permutation
symbol, this is $-\varepsilon_A^{\ B} D_B \BB = -2\Omega^a_A
\varepsilon_{apq} \Omega^q {\cal B}^p_{\ d} \Omega^d = 2 \Omega^a_A
\BB_a$, and we conclude that 
\begin{equation} 
\BB_A = -\frac{1}{2} \varepsilon_A^{\ B} D_B \BB, \qquad 
\BB := {\cal B}_{cd} \Omega^c \Omega^d. 
\label{4.2.3}
\end{equation} 
Acting on $\BB$ with two derivative operators and multiplying by the
Levi-Civita tensor gives $-\varepsilon_A^{\ C} D_B D_C \BB =
2\Omega^a_A \varepsilon_{apq} \Omega^p {\cal B}^q_{\ b} \Omega^b_B + 2 
\varepsilon_{AB} \BB$. Symmetrizing with respect to $A$ and $B$ and
consulting Table I yields $-(\varepsilon_A^{\ C} D_B 
+ \varepsilon_B^{\ C} D_A) D_C \BB = 2\Omega^a_A \Omega^b_B \BB_{ab}$, 
and we conclude that 
\begin{equation} 
\BB_{AB} = -\frac{1}{2} \bigl( \varepsilon_A^{\ C} D_B 
+ \varepsilon_B^{\ C} D_A \bigr) D_C \BB. 
\label{4.2.4}
\end{equation} 
This tensor is tracefree, because $\Omega^{AB} \BB_{AB} = -
\varepsilon^{BC} D_B D_C \BB = 0$ by virtue of the symmetry of $D_B
D_C \BB$ and the antisymmetry of the Levi-Civita tensor. 

\begin{table*}
\caption{Scalar and vectorial harmonics of degree $l=1$, labeled by 
the abstract index ${\sf m} = \{0, 1c, 1s\}$. The odd-parity vectorial
harmonics $X^{1{\sf m}}_A$ are not required, and the tensorial
harmonics $Y^{1{\sf m}}_{AB}$ and $X^{1{\sf m}}_{AB}$ vanish
identically.}   
\begin{ruledtabular} 
\begin{tabular}{cccc}
${\sf m}$ & $0$ & $1c$ & $1s$ \\
\hline
$Y^{1{\sf m}}$ & 
$\cos\theta$ & 
$\sin\theta\cos\phi$ &
$\sin\theta\sin\phi$ \\
$Y^{1{\sf m}}_\theta$ & 
$-\sin\theta$ & 
$\cos\theta\cos\phi$ & 
$\cos\theta\sin\phi$ \\ 
$Y^{1{\sf m}}_\phi$ & 
$0$ & 
$-\sin\theta\sin\phi$ &
$\sin\theta\cos\phi$ 
\end{tabular}
\end{ruledtabular}
\end{table*}   

\begin{table*}
\caption{Scalar, vectorial, and tensorial harmonics of degree $l=2$,
labeled by the abstract index ${\sf m} = \{0, 1c, 1s, 2c, 2s\}$.}    
\begin{ruledtabular} 
\begin{tabular}{cccccc}
${\sf m}$ & $0$ & $1c$ & $1s$ & $2c$ & $2s$ \\
\hline
$Y^{2{\sf m}}$ & $-(3\cos^2\theta-1)$ & 
$2\sin\theta\cos\theta\cos\phi$ & 
$2\sin\theta\cos\theta\sin\phi$ & 
$\sin^2\theta \cos 2\phi$ & 
$\sin^2\theta \sin 2\phi$ \\ 
$Y^{2{\sf m}}_\theta$ & 
$6\sin\theta\cos\theta$ & 
$2(2\cos^2\theta-1)\cos\phi$ & 
$2(2\cos^2\theta-1)\sin\phi$ & 
$2\sin\theta\cos\theta\cos 2\phi$ & 
$2\sin\theta\cos\theta\sin 2\phi$ \\ 
$Y^{2{\sf m}}_\phi$ & 
$0$ & 
$-2\sin\theta\cos\theta\sin\phi$ & 
$2\sin\theta\cos\theta\cos\phi$ & 
$-2\sin^2\theta\sin 2\phi$ & 
$2\sin^2\theta\cos 2\phi$ \\ 
$Y^{2{\sf m}}_{\theta\theta}$ & 
$-3\sin^2\theta$ & 
$-2\sin\theta\cos\theta\cos\phi$ & 
$-2\sin\theta\cos\theta\sin\phi$ & 
$(\cos^2\theta+1)\cos 2\phi$ & 
$(\cos^2\theta+1)\sin 2\phi$ \\ 
$Y^{2{\sf m}}_{\theta\phi}$ & 
$0$ & 
$2\sin^2\theta\sin\phi$ & 
$-2\sin^2\theta\cos\phi$ & 
$-2\sin\theta\cos\theta\sin 2\phi$ &
$2\sin\theta\cos\theta\cos 2\phi$ \\ 
$Y^{2{\sf m}}_{\phi\phi}$ & 
$3\sin^4\theta$ & 
$2\sin^3\theta\cos\theta\cos\phi$ & 
$2\sin^3\theta\cos\theta\sin\phi$ & 
$-\sin^2\theta(\cos^2\theta+1)\cos 2\phi$ & 
$-\sin^2\theta(\cos^2\theta+1)\sin 2\phi$ \\ 
$X^{2{\sf m}}_\theta$ & 
$0$ & 
$2\cos\theta\sin\phi$ & 
$-2\cos\theta\cos\phi$ & 
$2\sin\theta\sin 2\phi$ & 
$-2\sin\theta\cos 2\phi$ \\ 
$X^{2{\sf m}}_\phi$ & 
$6\sin^2\theta\cos\theta$ & 
$2\sin\theta(2\cos^2\theta-1)\cos\phi$ & 
$2\sin\theta(2\cos^2\theta-1)\sin\phi$ & 
$2\sin^2\theta\cos\theta\cos 2\phi$ & 
$2\sin^2\theta\cos\theta\sin 2\phi$ \\ 
$X^{2{\sf m}}_{\theta\theta}$ & 
$0$ & 
$-2\sin\theta\sin\phi$ & 
$2\sin\theta\cos\phi$ & 
$2\cos\theta\sin 2\phi$ & 
$-2\cos\theta\cos 2\phi$ \\ 
$X^{2{\sf m}}_{\theta\phi}$ & 
$-3\sin^3\theta$ & 
$-2\sin^2\theta\cos\theta\cos\phi$ & 
$-2\sin^2\theta\cos\theta\sin\phi$ & 
$\sin\theta(\cos^2\theta+1)\cos 2\phi$ & 
$\sin\theta(\cos^2\theta+1)\sin 2\phi$ \\ 
$X^{2{\sf m}}_{\phi\phi}$ & 
$0$ & 
$2\sin^3\theta\sin\phi$ & 
$-2\sin^3\theta\cos\phi$ & 
$-2\sin^2\theta\cos\theta\sin 2\phi$ &
$2\sin^2\theta\cos\theta\cos 2\phi$  
\end{tabular}
\end{ruledtabular}
\end{table*}   

Similar results can be obtained for the matter
potentials. Differentiating $\jj := j_a \Omega^a$ produces $D_A \jj =
j_a \Omega^a_A = j_c (\delta_a^{\ c} - \Omega_a \Omega^c) \Omega^a_A =
\Omega^a_A \jj_a$, and we conclude that 
\begin{equation} 
\jj_A = D_A \jj. 
\label{4.2.5}
\end{equation} 
Finally, 
\begin{equation} 
\SS_A = \frac{1}{2} D_A \SS 
\label{4.2.6}
\end{equation} 
follows after a calculation similar to the one leading to
Eq.~(\ref{4.2.1}). 

\subsection{Decomposition in spherical harmonics} 

The results obtained in the preceding subsection indicate that the  
tidal and matter potentials that appear in the metric of
Eqs.~(\ref{4.1.9})--(\ref{4.1.12}) can all be obtained by covariant
differentiation of the scalar potentials $\EE$, $\BB$, $\jj$, and
$\SS$. These are functions of the null coordinate $w$ and the
dependence on the angles $\theta^A$ appears in the factors
$\Omega^a(\theta^A)$. 

This angular dependence can be made more explicit by involving
spherical-harmonic functions. Let  
\begin{equation} 
Y^{1{\sf m}} = \bigl\{ Y^{1,0}, Y^{1,1c}, Y^{1,1s} \bigr\} 
\label{4.3.1}
\end{equation} 
be a set of real, unnormalized, spherical-harmonic functions of degree   
$l=1$. And let 
\begin{equation} 
Y^{2{\sf m}} = \bigl\{ Y^{2,0}, Y^{2,1c}, Y^{2,1s}, Y^{2,2c}, 
              Y^{2,2s} \bigr\}  
\label{4.3.2}
\end{equation} 
be a set of real, unnormalized, spherical-harmonic functions of degree  
$l=2$. The abstract index $\sf m$ describes the dependence of the
spherical-harmonics on the angle $\phi$; the numerical part of the
label refers to the azimuthal index $m$, and the letter indicates
whether the function is proportional to $\cos(m\phi)$ or
$\sin(m\phi)$. Explicit expressions are listed in Tables II and III.  

We decompose the scalar potentials according to 
\begin{eqnarray} 
\EE(w,\theta^A) &=& 
\sum_{\sf m} \EE_{\sf m}(w) Y^{2{\sf m}}(\theta^A), 
\label{4.3.3} \\ 
\BB(w,\theta^A) &=& 
\sum_{\sf m} \BB_{\sf m}(w) Y^{2{\sf m}}(\theta^A), 
\label{4.3.4} \\ 
\jj(w,\theta^A) &=& 
\sum_{\sf m} \jj_{\sf m}(w) Y^{1{\sf m}}(\theta^A), 
\label{4.3.5} \\ 
\SS(w,\theta^A) &=& 
\sum_{\sf m} \SS_{\sf m}(w) Y^{2{\sf m}}(\theta^A), 
\label{4.3.6} 
\end{eqnarray} 
in terms of their harmonic components $\EE_{\sf m}$, $\BB_{\sf m}$, 
$\jj_{\sf m}$, and $\SS_{\sf m}$. These are in a one-to-one
correspondence with the frame tensors ${\cal E}_{ab}(w)$, 
${\cal B}_{ab}(w)$, $j_{a}(w)$, and $S_{ab}(w)$; the relationships are 
displayed in Table IV.  

The derivatives of the scalar potentials will be expressed in terms of
derivatives of the spherical-harmonic functions. The vectorial
harmonics are 
\begin{equation} 
Y^{l{\sf m}}_A = D_A Y^{l{\sf m}}, \qquad 
X^{l{\sf m}}_A = -\varepsilon_A^{\ B} D_B Y^{l{\sf m}}, 
\label{4.3.7}
\end{equation} 
and the tensorial harmonics are 
\begin{equation} 
Y^{l{\sf m}} \Omega_{AB}, \qquad 
Y^{l{\sf m}}_{AB} = \bigl[ D_A D_B + {\textstyle \frac{1}{2}} l(l+1)
\Omega_{AB} \bigr] Y^{l{\sf m}}
\label{4.3.8} 
\end{equation}
and 
\begin{equation} 
X^{l{\sf m}}_{AB} = -\frac{1}{2} \bigl( \varepsilon_A^{\ C} D_B  
+ \varepsilon_B^{\ C} D_A \bigr) D_C Y^{l{\sf m}}. 
\label{4.3.9}
\end{equation} 
Apart from notation and normalization, these definitions agree with
those of Regge and Wheeler \cite{regge-wheeler:57}. We note that the
tensorial harmonics $Y^{l{\sf m}}_{AB}$ and $X^{l{\sf m}}_{AB}$ are
symmetric and tracefree. 

The decompositions of the vectorial and tensorial potentials in terms 
of vectorial and tensorial harmonics are displayed in Table IV. They
are obtained by substituting Eqs.~(\ref{4.3.3})--(\ref{4.3.6}) into 
Eqs.~(\ref{4.2.1})--(\ref{4.2.6}).    

The most explicit form for the metric tensor is obtained after
substituting the spherical-harmonic decompositions of Table IV, along
with the spherical-harmonic functions listed in Tables II and III,
into Eqs.~(\ref{4.1.9})--(\ref{4.1.12}). This leads to long
expressions, but in practical applications it may happen that only a
few frame components among ${\cal E}_{ab}$, ${\cal B}_{ab}$, $\rho$,
$j_a$, $S_{ab}$, and $T$ are nonzero; in such cases only a few
harmonic components among $\EE_{\sf m}$, $\BB_{\sf m}$, $\jj_{\sf m}$,
and $\SS_{\sf m}$ will contribute to the metric, and the expressions
will simplify. We shall encounter such cases in the next section. 

\begin{table*}
\caption{Spherical-harmonic decomposition of the tidal and matter
potentials. In the first part of the table we list the definitions of 
$\jj_{\sf m}(w)$, $\EE_{\sf m}(w)$, $\BB_{\sf m}(w)$, and 
$\SS_{\sf m}(w)$ in terms of $j_a(w)$, ${\cal E}_{ab}(w)$, 
${\cal B}_{ab}(w)$, and $S_{ab}(w)$. In the second part of the table
we display the spherical-harmonic decompositions of the tidal and
matter potentials.}  
\begin{ruledtabular} 
\begin{tabular}{llll}
$\jj_0 = j_3$ & 
$\EE_0 = \frac{1}{2}({\cal E}_{11} + {\cal E}_{22})$ & 
$\BB_0 = \frac{1}{2}({\cal B}_{11} + {\cal B}_{22})$ & 
$\SS_0 = \frac{1}{2}(S_{11} + S_{22})$ \\ 
$\jj_{1c} = j_1$ & 
$\EE_{1c} = {\cal E}_{13}$ &
$\BB_{1c} = {\cal B}_{13}$ & 
$\SS_{1c} = S_{13}$ \\ 
$\jj_{1s} = j_2$ &
$\EE_{1s} = {\cal E}_{23}$ &
$\BB_{1s} = {\cal B}_{23}$ & 
$\SS_{1s} = S_{23}$ \\
 & 
$\EE_{2c} = \frac{1}{2} ({\cal E}_{11} - {\cal E}_{22})$ & 
$\BB_{2c} = \frac{1}{2} ({\cal B}_{11} - {\cal B}_{22})$ & 
$\SS_{2c} = \frac{1}{2} (S_{11} - S_{22})$ \\
 & 
$\EE_{2s} = {\cal E}_{12}$ &
$\BB_{2s} = {\cal B}_{12}$ &
$\SS_{2s} = S_{12}$ \\
\hline 
$\jj = \sum_{\sf m} \jj_{\sf m} Y^{1 {\sf m}}$ & 
$\EE = \sum_{\sf m} \EE_{\sf m} Y^{2 {\sf m}}$ & 
 & 
$\SS = \sum_{\sf m} \SS_{\sf m} Y^{2 {\sf m}}$ \\ 
$\jj_A = \sum_{\sf m} \jj_{\sf m} Y_A^{1 {\sf m}}$ & 
$\EE_A = \frac{1}{2} \sum_{\sf m} \EE_{\sf m} Y^{2 {\sf m}}_{A}$ & 
$\BB_A = \frac{1}{2} \sum_{\sf m} \BB_{\sf m} X^{2 {\sf m}}_{A}$ & 
$\SS_A = \frac{1}{2} \sum_{\sf m} \SS_{\sf m} Y^{2 {\sf m}}_{A}$ \\ 
 & 
$\EE_{AB} = \sum_{\sf m} \EE_{\sf m} Y^{2 {\sf m}}_{AB}$ & 
$\BB_{AB} = \sum_{\sf m} \BB_{\sf m} X^{2 {\sf m}}_{AB}$ & 
\end{tabular}
\end{ruledtabular}
\end{table*}   

\section{Applications} 

\subsection{Comoving observer in a spatially-flat cosmology} 

To illustrate how the formalism works we first consider the world line
of a comoving observer in a cosmological spacetime. The global metric
is  
\begin{equation} 
ds^2 = -dt^2 + a^2(t)\bigl( dx^2 + dy^2 + dz^2 \bigr), 
\label{5.1.1}
\end{equation}
where $a(t)$ is an arbitrary scale factor; for simplicity we take the
cosmology to be spatially flat. This application was already presented  
in Ref.~\cite{poisson:04a}, but we generalize it here from the retarded
coordinates considered there to light-cone coordinates of both types
(retarded and advanced). Furthermore, the decomposition of the 
energy-momentum tensor into irreducible parts was not accomplished in  
the earlier paper, and this decomposition adds insight to our earlier
results.    

Without loss of generality we take our observer to be at the spatial
origin of the global coordinate system ($x=y=z=0$), and her velocity
vector is given by 
\begin{equation}  
u^\mu = (1,0,0,0)
\label{5.1.2}
\end{equation} 
in the ordering $(t,x,y,z)$ of the cosmological coordinates. This
vector satisfies the geodesic equation, and $t$ is proper time for the  
observer. We wish to transform the metric of Eq.~(\ref{5.1.1}) to
light-cone coordinates $(w,r,\theta^A)$ centered on the world line of
this observer.      

To do so we must first construct a triad of orthonormal spatial
vectors $\base{\mu}{a}$. A simple choice is
\begin{eqnarray} 
\base{\mu}{1} &=& (0,a^{-1},0,0), 
\nonumber \\  
\base{\mu}{2} &=& (0,0,a^{-1},0), 
\label{5.1.3} \\ 
\base{\mu}{3} &=& (0,0,0,a^{-1}); 
\nonumber 
\end{eqnarray}
these vectors are all parallel transported on $\gamma$. 

According to Eq.~(\ref{3.2.2}) and the fact that the Weyl tensor of
the spacetime vanishes, we have 
\begin{equation}
{\cal E}_{ab} = {\cal B}_{ab} = 0. 
\label{5.1.4}
\end{equation} 
And according to Eq.~(\ref{3.2.3}) and a simple computation, we have 
$j_a = S_{ab} = 0$ and 
\begin{equation} 
\rho = \frac{3}{8\pi} (\dot{a}/a)^2, \quad 
T = -\frac{3}{8\pi} \bigl[ 2\ddot{a}/a 
+ (\dot{a}/a)^2 \bigr]. 
\label{5.1.5}
\end{equation} 
Here the scale factor is expressed in terms of $w \equiv 
\mbox{[proper time on $\gamma$]}$ by simply making the functional
substitution $a(t) \to a(w)$; overdots indicate differentiation with
respect to $w$. Recall that $\rho$ is the mass-energy density measured
by the observer, and that $\frac{1}{3} T$ is the measured pressure of
the cosmological fluid.   

The vanishing of ${\cal E}_{ab}$, ${\cal B}_{ab}$, $j_a$, and $S_{ab}$ 
implies that the metric is spherically symmetric around $\gamma$
(this does not come as a surprise). After substituting
Eqs.~(\ref{5.1.4}) and (\ref{5.1.5}) into
Eqs.~(\ref{4.1.9})--(\ref{4.1.12}), a short calculation reveals that
the metric components are given by   
\begin{eqnarray} 
\hspace*{-15pt} 
g_{ww} &=& -1 + r^2 (\ddot{a}/a) + O(r^3),
\label{5.1.6} \\ 
\hspace*{-15pt} 
g_{wr} &=& \eta, 
\label{5.1.7} \\ 
\hspace*{-15pt} 
g_{wA} &=& O(r^4), 
\label{5.1.8} \\ 
\hspace*{-15pt} 
g_{AB} &=& r^2 \Omega_{AB} \biggl\{ 1 + \frac{1}{3} r^2 \bigl[
\ddot{a}/a - (\dot{a}/a)^2 \bigr] + O(r^3) \biggr\}. 
\label{5.1.9} 
\end{eqnarray}
We recall that the scale factor and its derivatives are functions
of the null coordinate $w$. When the scale factor behaves as a power
law, $a(t) \propto t^\alpha$ with $\alpha$ a constant, we have
$\ddot{a}/a = -\alpha(1-\alpha)/w^2$ and $\ddot{a}/a - (\dot{a}/a)^2 = 
-\alpha/w^2$. When instead the scale factor behaves as an exponential,
$a(t) \propto e^{H t}$ with $H$ a constant, we have $\ddot{a}/a = H^2$
and $\ddot{a}/a - (\dot{a}/a)^2 = 0$.  

\subsection{Static observer in Melvin's magnetic universe} 

Melvin's magnetic universe \cite{melvin:64, melvin:65, thorne:65} is a
static, cylindrically symmetric spacetime that is filled with a
magnetic field held together by gravity. The exact solution to the
Einstein-Maxwell equations that describes this situation consists of
the metric  
\begin{equation} 
ds^2 = \Lambda^2 \bigl( -dt^2 + d\bar{\rho}^2 + dz^2 \bigr) 
+ \Lambda^{-2} \bar{\rho}^2\, d\varphi^2 
\label{5.2.1} 
\end{equation}
and the vector potential 
\begin{equation} 
A^\alpha = \frac{1}{2} B \Lambda \varphi^\alpha, 
\label{5.2.2}
\end{equation}
where $\varphi^\alpha = \partial x^\alpha/d\varphi$ is the spacetime's 
azimuthal Killing vector. We have introduced 
\begin{equation} 
\Lambda := 1 + \frac{1}{4} B^2 \bar{\rho}^2, 
\label{5.2.3}
\end{equation} 
and the constant $B$ measures the strength of the magnetic field. The
metric and the vector potential are expressed in cylindrical
coordinates $(t,\bar{\rho},z,\varphi)$. 

The metric of Eq.~(\ref{5.2.1}) can be decomposed in terms of a tetrad 
of orthonormal vectors. We introduce a ``Cartesian'' frame described
by  
\begin{eqnarray} 
\base{\alpha}{0} &:=& \bigl( \Lambda^{-1},0,0,0 \bigr), 
\label{5.2.4} \\ 
\base{\alpha}{1} &:=& \bigl( 0,\Lambda^{-1}\cos\varphi,0,
-\Lambda\bar{\rho}^{-1}\sin\varphi \bigr), 
\label{5.2.5} \\ 
\base{\alpha}{2} &:=& \bigl( 0,\Lambda^{-1}\sin\varphi,0,
\Lambda\bar{\rho}^{-1}\cos\varphi \bigr), 
\label{5.2.6} \\ 
\base{\alpha}{3} &:=& \bigl( 0,0,\Lambda^{-1},0 \bigr).  
\label{5.2.7}
\end{eqnarray} 
It is easy to check that the inverse metric can be expressed as 
$g^{\alpha\beta} = -\base{\alpha}{0}\base{\beta}{0} 
+ \base{\alpha}{1}\base{\beta}{1} + \base{\alpha}{2}\base{\beta}{2}
+ \base{\alpha}{3}\base{\beta}{3}$. It is also easy to check that in
this tetrad, the electromagnetic field tensor has 
\begin{equation}
B_3 := F_{12} := F_{\alpha\beta} \base{\alpha}{1} \base{\beta}{2} 
= \frac{B}{\Lambda^2} 
\label{5.2.8}
\end{equation} 
as its only nonvanishing component. 

We wish to consider a static observer in Melvin's magnetic
universe. To ensure that this observer moves on a world line $\gamma$
that is a geodesic of the spacetime, we place him on the axis of
symmetry at $\bar{\rho} = 0$. The observer has a velocity vector
given by $u^\alpha = \base{\alpha}{0}(\bar{\rho} = 0)$, and
$\base{\alpha}{a}(\bar{\rho}=0)$ is a triad of parallel-transported
vectors on $\gamma$. 

A straightforward computation reveals that the metric of
Eq.~(\ref{5.2.1}) comes with a Weyl tensor whose nonvanishing electric
components are 
\begin{equation} 
{\cal E}_{11} = {\cal E}_{22} = \frac{1}{2} B^2, \qquad 
{\cal E}_{33} = -B^2. 
\label{5.2.9}
\end{equation} 
The magnetic part of the Weyl tensor vanishes: ${\cal B}_{ab} 
= 0$. A computation of the energy-momentum tensor (either from the
metric or from the electromagnetic field tensor) reveals that  
\begin{equation} 
\rho = \frac{B^2}{8\pi}, 
\label{5.2.10} 
\end{equation}
\begin{equation} 
S_{11} = S_{22} = \frac{B^2}{12\pi}, \qquad 
S_{33} = -\frac{B^2}{6\pi}, 
\label{5.2.11}
\end{equation}
and 
\begin{equation} 
T = \frac{B^2}{8\pi}, 
\label{5.2.12} 
\end{equation} 
while $j_a = 0$. (Recall that $T := \delta^{ab} T_{ab}$ is the
three-dimensional trace of the energy-momentum tensor; the
four-dimensional trace is $\mbox{}^4 T = T - \rho$, and it vanishes by
virtue of the conformal invariance of Maxwell's equations.) These
relations imply that $\EE_0 = \frac{1}{2} B^2$ and $\SS_0 
= B^2/(12\pi)$ are the only nonvanishing harmonic components of the
tidal and matter potentials.  

Making the substitutions from Eqs.~(\ref{5.2.9})--(\ref{5.2.12}),
Table III, and Table IV into Eqs.~(\ref{4.1.9})--(\ref{4.1.12}), we
find that the metric components in light-cone coordinates are 
\begin{eqnarray} 
g_{ww} &=& -1 - \frac{1}{2} B^2 r^2 \sin^2\theta + O(r^3), 
\label{5.2.13} \\ 
g_{wr} &=& \eta, 
\label{5.2.14} \\ 
g_{w\theta} &=& -\frac{1}{3} \eta B^2 r^3 \sin\theta\cos\theta 
+ O(r^4), 
\label{5.2.15} \\ 
g_{\theta\theta} &=& r^2 + \frac{1}{6} B^2 r^4 \sin^2\theta 
+ O(r^5), 
\label{5.2.16} \\ 
g_{\phi\phi} &=& r^2\sin^2\theta - \frac{5}{6} B^2 r^4 \sin^4\theta 
+ O(r^5). 
\label{5.2.17} 
\end{eqnarray} 
As expected, the metric is axially symmetric, but the full cylindrical  
symmetry of the spacetime is not revealed by the light-cone
coordinates.   
    
\begin{acknowledgments} 
This work was supported by the Natural Sciences and Engineering
Research Council of Canada.     
\end{acknowledgments} 

\bibliography{../bib/master}
\end{document}